%% file: main.tex
\documentclass[acmsmall,screen]{acmart}
\AtBeginDocument{%
  }

\usepackage{pifont}
\usepackage{hyperref}
\usepackage{multirow}
\usepackage{graphicx}
\usepackage[most]{tcolorbox}

\tcbset{
  RQBox/.style={
    colback=gray!20,    
    colframe=gray!80,   
    fonttitle=\bfseries,
    coltitle=black,
    boxrule=0.5pt,
    arc=4pt,
    left=2mm,
    right=2mm,
    top=1mm,
    bottom=1mm
  }
}

\setcopyright{acmlicensed}
\copyrightyear{2018}
\acmYear{2018}
\acmDOI{XXXXXXX.XXXXXXX}
\acmConference[Conference acronym 'XX]{Make sure to enter the correct
  conference title from your rights confirmation email}{June 03--05,
  2018}{Woodstock, NY}
\acmISBN{978-1-4503-XXXX-X/2018/06}

\usepackage{enumitem}

\input{listing_style}

\usepackage{xspace}
\newcommand{\toolname}[0]{\textsc{MuMuTestUp}\xspace}
\newcommand{\dataset}[0]{\textsc{Prbench}\xspace}
\newcommand{\testupdater}[0]{\textsc{TestUpdater}\xspace}
\newcommand{\target}[0]{\textsc{TaRGET}\xspace}
\newcommand{\synter}[0]{\textsc{Synter}\xspace}
\newcommand{\reaccept}[0]{\textsc{ReAccept}\xspace}
\newcommand{\tarbench}[0]{\textsc{TaRBench}\xspace}

\newcommand{\inputProcessing}[0]{Input Preprocessing\xspace}
\newcommand{\testUpdate}[0]{Test Update\xspace}
\newcommand{\coordinator}[0]{Coordinator\xspace}
\newcommand{\errorAnalysis}[0]{Error Analysis\xspace}
\newcommand{\coverageAnalysis}[0]{Coverage Analysis\xspace}
\newcommand{\mutationAnalysis}[0]{Mutation Analysis\xspace}
\newcommand{\retrievaler}[0]{Semantic Retrieval\xspace}

\author{Dawei Tian}
\affiliation{%
  \institution{Harbin Institute of Technology}
  \city{Harbin}
  \country{China}}
\email{davidtian@stu.hit.edu.cn}
\orcid{0009-0005-6476-790X}

\author{Jiakun Liu*}
\affiliation{%
  \institution{Harbin Institute of Technology}
  \city{Harbin}
  \country{China}}
\email{jiakunliu@hit.edu.cn}

\author{Yun Peng}
\affiliation{%
  \institution{The Chinese University of Hong Kong}
  \city{Hong Kong}
  \country{China}}
\email{ypeng@cse.cuhk.edu.hk}

\author{Yichen Zhang}
\affiliation{%
  \institution{Harbin Institute of Technology}
  \city{Harbin}
  \country{China}}
\email{yichenzhang683@gmail.com}

\author{Jianlei Chi}
\affiliation{%
  \institution{Xidian University}
  \city{Xi'an}
  \country{China}}
\email{chijianlei@gmail.com}

\author{Jun Sun}
\affiliation{%
  \institution{Singapore Management University}
  \country{Singapore}}
\email{junsun@smu.edu.sg}

\author{Xiaohong Su}
\affiliation{%
  \institution{Harbin Institute of Technology}
  \city{Harbin}
  \country{China}}
\email{sxh@hit.edu.cn}

\begin{document}

\title{$MuMuTestUp$: Mutation-based Multi-Agent Test Case Update}

\begin{abstract}
Modern software systems evolve rapidly under continuous integration and deployment (CI/CD) practices, in which tests act as critical gatekeepers of software quality. However, when substantial code changes are introduced, existing test cases may become obsolete, leading to compilation failures, erroneous test behaviors, or inadequate regression coverage. Such issues can disrupt CI/CD pipelines, degrade development productivity, and ultimately undermine overall software quality. Many efforts are devoted to designing automatic test case update methods to address these issues. The most recent approaches rely on large language models (LLMs) to iteratively refine test cases using execution feedback from compilation errors or coverage reports, and on context retrieved via exact-matching approaches. They also prioritize test executability and line coverage to quickly build executable, correct test cases from the original broken test cases. Despite their correctness, current approaches face three limitations: (1) they focus on executabilty but overlook the adequacy of test assertions, which lowers the capability of test cases to detect faults; (2) they utilize only coarse line coverage singals instead of specific information about uncovered lines and branches; (3) they use exact-matching context retrieval approaches, which fails to provide accurate context given potential hallucinated queries from LLMs.

To address these challenges, we propose \toolname, a \textbf{Mu}tation-guided, \textbf{Mu}lti-agent framework for automated \textbf{test} case \textbf{up}dating. \toolname integrates three specialized agents: (1) a \mutationAnalysis agent that leverages surviving mutants as indicators of weak or missing test assertions and generates individual repair instructions to strengthen or synthesize assertions for each surviving mutant, (2) a \coverageAnalysis agent generates individual repair instructions for each uncovered line, uncovered branch rather than exposing raw coverage signals to the LLM, and (3) a \retrievaler agent that uses semantic-similarity search to handle unavailable or hallucinated symbols. Additionally, we construct \dataset, a pull-request–level dataset of 571 samples from 10 open-source Java projects that considered cross-commit update scenarios, validated through three rounds of execution following prior studies to detect outdated tests. We evaluate \toolname against state-of-the-art baselines using both open-source and closed-source LLMs (Deepseek-V3.2 and GPT-4.1). With GPT-4.1, \toolname achieves a line coverage of 88.94\%, branch coverage of 63.36\%, and mutation score of 72.39\%, outperforming the best baseline by 5.33\%, 19.93\%, and 16.66\%, respectively. 


\end{abstract}

\maketitle

\input{sections/intro}

\input{sections/problem_definition}

\input{sections/motivating}

\input{sections/approach}

\input{sections/data}

\input{sections/experiment}

\input{sections/result}

\input{sections/discussion}

\input{sections/related_work}

\section{Conclusion}

In this paper, we presented \toolname, a mutation-guided multi-agent framework for automated test case updating. By integrating fine-grained execution feedback, coverage analysis, mutation analysis, and semantic retrieval, \toolname goes beyond restoring test executability and systematically improves both functional coverage and assertion robustness. Extensive experiments on a large-scale benchmark of 571 real-world test update instances show that \toolname consistently outperforms the best baseline by 5.33\%, 19.93\%, and 16.66\% in terms of line coverage, branch coverage, and mutation score, respectively, and remains robust across different backend LLMs. These results indicate that effective test case updating requires jointly considering executability, coverage completeness, and oracle strength. An important direction for future work is to extend \toolname to additional programming languages and build systems.

\section*{Data Availability}
Our code, scripts, prompts, descriptions of the tools used in the \toolname, experimental results, and dataset are publicly available at: \url{https://github.com/crazyTang-cloud/MuMuTestUp.git}.


\bibliographystyle{ACM-Reference-Format}
\bibliography{reference}

\end{document}

%% file: listing_style.tex
\usepackage{listings}

\definecolor{codegreen}{rgb}{0,0.6,0}
\definecolor{codegray}{rgb}{0.5,0.5,0.5}
\definecolor{codepurple}{rgb}{0.58,0,0.82}
\definecolor{backcolour}{rgb}{0.95,0.95,0.92}
\definecolor{lightgreen}{rgb}{0,0.4,0}
\definecolor{lightred}{rgb}{0.4,0,0}

\definecolor{mygray}{gray}{.9}

\lstdefinelanguage{JavaScript}{
  keywords={typeof, new, true, false, catch, function, return, null, catch, switch, var, if, in, while, do, else, case, break, export, static, const, let, class, void, async, await, interface, extends, public, this},
  ndkeywords={class, export, boolean, throw, implements, import, this},
  ndkeywordstyle=\color{darkgray}\bfseries,
  identifierstyle=\color{black},
  sensitive=false,
  frame=lines,
  comment=[l]{//},
  morecomment=[s]{/*}{*/},
  morestring=[b]',
  morestring=[b]"
}

%% file: sections/intro.tex
\section{Introduction} \label{sec:intro}
With software development pipelines such as continuous integration and deployment (CI/CD), modern software systems evolve fastly~\cite{Runeson_jsm2003,Rajlich_slc_2000,Skoglund_icsm2004}. Test cases play important roles in modern software development process since they validate and guarantee the correct functionality of software \cite{test_important_verma_2023ijmem,test_important_Petunova_2017itms,test_important_Jia_2023jiem}. However, test cases can become outdated and broken with the rapid changes of requirements and code, which further leads to compilation failures, runtime errors, or missed regressions, threatening the software reliability and the efficiency of software development \cite{Labuschagne_regression_test_fse2017,Vahabzadeh_test_empricial_icsme2015}.
To maintain software quality \cite{Olan_csc2003,Duvall_book_awp2007,Herzig_icse2015}, test cases must therefore be continuously co-evolved with production code to detect regressions as early as possible.

Manually updating test cases to adapt code changes is time-consuming and error-prone \cite{Kasurinen2010SoftwareTA,Widder_CI_FSE2019}, especially when test cases and code are often maintained by different teams in practice.
Furthermore, software evolution is often complex, with individual updates spanning dozens or even hundreds of code changes. This makes comprehensive manual reasoning about test updates infeasible. Therefore, automating the evolution of outdated test cases is crucial for modern software development and maintanance.


Early studies on automated test case update rely on program analysis techniques and heuristic rules \cite{Daniel_test_update_issta2010,Daniel_test_update_tool_icse2011,Xu_test_update_apsec2014,Mehdi_test_update_icst2012}, and they are inherently limited in handling the diverse update scenarios encountered in practice.
With advances in machine learning, more recent work incorporates pre-trained language models for test case update \cite{CEPROT,TARGET}.
For example, \target \cite{TARGET} employs CodeT5 to update outdated tests but is restricted to single-hunk changes.
More recently, large language model (LLM)-based approaches are proposed for automatic test case update.
For example, Chi et al. \cite{REACCEPT} introduce an LLM-based approach augmented with execution feedback and retrieval, Xu et al. \cite{CommitUp} infer change intent to guide test updates, and Zhang et al. \cite{TestUpdater} proposed a post-hoc retrieval strategy to supply failure-related context.
Despite the advances, existing research on automatic test case update has the following limitations.

\textbf{\ding{202} Insufficient focus on test assertions.} 
With prioritization on executability, existing approaches tend to generate weak assertions so the updated test cases can be successfully executed. However, test cases with weak assertions usually have low mutation scores \cite{Panichella_dynamosa_tse2018,DAKHEL2024107468_ist,Harman_ACH_icse2025}, indicating that they are not capable to discriminate between correct and abnormal behavior of code. Instead, they may always succeed in any programs. This further threatens the quality assurance capability of the entire test suite. 

\textbf{\ding{203} Inappropriate line coverage feedback.} 
Existing approaches use line coverage as feedback, focusing on covering lines \cite{CommitUp} or newly introduced code \cite{REACCEPT,TestUpdater}. However, line coverage measures only the syntactic execution of code and cannot guarantee that all logical branches are covered.  Without branch coverage guidance, updated test cases may fail to validate important conditional logic, and thus they can reduce the effectiveness of regression testing \cite{Yang2024EnhancingLT,HITS,Fraser_evosuite_fse2011}.

\textbf{\ding{204} Limited context retrieval based on exact matching.} Existing 
studies \cite{REACCEPT,CommitUp,CEPROT,SITAR} underestimated the importance of context when updating test cases. Context is essential in test case update as it provides necessary information about related code elements and repository knowledge. To date, only \synter \cite{SYNTER} and \testupdater \cite{TestUpdater} used language-server–based retrieval with exact matching. 
However, context retrieval methods based on exact matching can hardly handle the queries that contain hallucinated symbols generated by LLMs, limiting the robustness of test case updates in practice. 

To address these limitations, we propose \toolname, a \textbf{Mu}tation-guided \textbf{Mu}lti-agent-based \textbf{Test} Case \textbf{Up}date framework.
Specifically, to address Limitation \ding{202}, we design a \mutationAnalysis agent, inspired by the mutation testing techniques \cite{Fraser_mutation_test_tse2012,Fraser_mosa_tse2013,Hamlet_mutation_test_tse1977,DeMillo_mutation_test_Computer1978,Fraser_evosuite_fse2011,Lukasczyk_pynguin_icse_2022} that inject faults (mutants) into code to guide the generation of effective test assertions. It generates repair instructions for each surviving mutant, enabling the LLM to strengthen or generate strong assertions that specifically detect these mutants. To address Limitation \ding{203}, we design a \coverageAnalysis agent that generates repair instructions for each uncovered line and branch. This allows the LLM to update tests with precise guidance, avoiding blind exploration and effectively improving coverage. To address Limitation \ding{204}, we design a \retrievaler agent that retrieves necessary contextual symbols to guide test updates. To handle hallucinated or otherwise unavailable symbols generated by the LLM, the agent employs a vector database with semantic similarity matching, enabling robust retrieval.


%

%

%

%

Existing datasets for test case update include only samples where the focal method has changed and are often constructed at a single-commit level \cite{SITAR,CEPROT,REACCEPT,TestUpdater,CommitUp,TARGET}. In practice, test cases can become outdated due to changes in other methods or across multiple commits. Therefore, datasets covering both focal-method–changed and –unchanged cases across multiple commits are essential for realistically evaluating test case update techniques. To evaluate our approach, we construct a new pull-request-level test case dataset \dataset comprising 571 samples from 10 open-source Java projects on GitHub. Specifically, we mined outdated test cases at the pull request (PR) level and applied a three-round execution to collect samples containing outdated test cases.

We compare our approach against three state-of-the-art (SOTA) baselines, including one CLM-based approach, \target \cite{TARGET}, and two LLM-based approaches, \reaccept \cite{REACCEPT} and \testupdater \cite{TestUpdater}. We further evaluate the performance of our approach on both open-source and closed-source LLMs, specifically GPT-4.1 and Deepseek-V3.2, and compare compilation pass rate, test pass rate, line coverage, branch coverage, and mutation score across different approaches.
When using GPT-4.1, our approach achieves a line coverage of 88.94\%, a branch coverage of 63.36\%, and a mutation score of 72.39\%. These results outperform the best baseline, \testupdater by 5.33\%, 19.93\%, and 16.66\%, respectively.
When using Deepseek-V3.2, our approach likewise demonstrates substantial improvements across all metrics.  

The main contributions of this paper are as follows:

\begin{itemize}[wide=0pt]
    \item To the best of our knowledge, we are the first to systematically integrate assertion guidance via mutation analysis into automated test updating, bridging a longstanding gap where existing research focused almost exclusively on structural coverage guidance. 
    \item We design a multi-agent framework that separates error analysis, coverage analysis, mutation analysis, and contextual retrieval into specialized agents, enabling coordinated optimization of coverage, assertion robustness, and update correctness. 
    \item We propose a novel semantic retrieval strategy, improving retrieval effectiveness and preserving the overall stability of the system.
    \item We construct the first dataset \dataset for pull-request-level test case updating and demonstrate, based on this dataset, that our approach achieves significant improvements over existing approaches.
\end{itemize}

%% file: sections/problem_definition.tex
\section{Problem Definition}\label{sec_definition}

In the context of software evolution, we consider the problem of automatically updating test cases to maintain their effectiveness and behavioral adequacy. Let $T_{before}$ denote the test case before the code change (at any granularity, e.g., code edit, commit, or pull request). Similarly, $F_{before}$ and $F_{after}$ represent the method under test (focal method) before and after the code change, respectively.In practice, even if the target method $F_{before}$ may remain unchanged, modifications in the call chain of the test case can still cause the corresponding test case $T_{before}$ to be outdated.
For example, a functional change to a method, even if that method is not $F_{before}$ itself but is called by $F_{before}$, can still cause $T_{before}$ to fail and report an error.
If the target method remains unchanged, then $F_{before} = F_{after}$.
The changes in the production code are represented as a series of diff hunks, $\text{diff}_1, \dots, \text{diff}_n$, and $T_{update}$ represents the updated test suite.

The input for this task includes the original test case $T_{before}$ and its context within the test class (i.e., non-test methods $NT_{before}$ \footnote{For example, methods annotated with @Before or @After in JUnit, and private helper methods.} and class or member variables $V_{before}$), the diff hunks $\text{diff}_1, \dots, \text{diff}_n$, and the focal methods $F_{before}$ and $F_{after}$ (if $F_{before} \neq F_{after}$; otherwise $F_{before}$). The goal is to generate an updated test case $T_{update}$ that compiles and executes successfully while maintaining sufficient robustness and assertion coverage.

%% file: sections/motivating.tex
\section{Motivation Example}

This section presents two motivating examples that highlight the key limitations of existing approaches. We systematically applied state-of-the-art methods, including \reaccept and \testupdater, as well as our approach, \toolname, to these examples. As \testupdater failed to generate compilable test cases for either example, we report the results of \reaccept and \toolname for clarity.


\begin{lstlisting}[language=JavaScript, xleftmargin=2em, frame=none, numbers=left, captionpos=b, caption=A example of Limitation \ding{202}. This example illustrates that\, to restore test executability\, prior approaches may remove the original assertion logic\, thereby weakening the test oracle., label=fig:motivation_example2]
# Outdated test
public void testSetMissingPropertyTypeFromStringValue() {
    assertEquals(this.propertyRegister.setPropertyTypeFromStringValue(
        dataObjectFactory.getPropertyIdValue("P10", this.siteIri),
        dataObjectFactory.getStringValue("http://musicbrainz.org/$1/artist")), 
        "http://www.wikidata.org/ontology#propertyTypeString" //outdated
    );
}

# Updated by ReAccept
public void testSetMissingPropertyTypeFromStringValue() {
    assertEquals(this.propertyRegister.setPropertyTypeFromStringValue(
        dataObjectFactory.getPropertyIdValue("P10", this.siteIri),
        dataObjectFactory.getStringValue("http://musicbrainz.org/$1/artist")));
+    // After calling the method, we might need to verify the property type was set correctly
+    // This might require additional assertions checking the state of propertyRegister
}

# Updated by MuMuTestUp
public void testSetMissingPropertyTypeFromStringValue() {
    assertEquals(
+        "http://wikiba.se/ontology#String",
        this.propertyRegister.setPropertyTypeFromStringValue(
            dataObjectFactory.getPropertyIdValue("P10", this.siteIri),
            dataObjectFactory.getStringValue("http://musicbrainz.org/$1/artist")));
}
\end{lstlisting}

The example shown in Listing~\ref{fig:motivation_example2} illustrates Limitation~\ding{202} of existing approaches.\footnote{From pull request 223 in the Wikidata-Toolkit project: \url{https://github.com/Wikidata-Toolkit/Wikidata-Toolkit/pull/223}.}
In this example, the test case \verb|testSetMissingPropertyTypeFromStringValue()| contains an assertion that becomes outdated after the pull request is merged. While the assertion correctly captured the expected behavior prior to the change, it fails under the updated production code. Correctly updating the test requires revising the expected value, specifically by replacing \texttt{http://www.wikidata.org/ontology/\allowbreak\#propertyTypeString} with \texttt{http://wikiba.se/ontology\#String}.
However, to ensure the test passes, \reaccept removes the original assertion logic entirely, thereby weakening the test oracle. In contrast, \toolname leverages mutation-testing–based assertion feedback to preserve the original assertion structure and correctly repair the assertion to reflect the updated behavior.

\begin{lstlisting}[language=JavaScript, xleftmargin=2em, frame=none, numbers=left, captionpos=b, caption=An example of Limitation \ding{203}. This example shows that some SOTA approaches restore compilability by removing faulty code\, while overlooking whether the code was successfully executed before the change., label=fig:motivation_example1]
# Outdated test
public void testInjectHtml() {
    ... //five behavioral checks omit in the paper due to the page limitation
    // overriden value
    System.getProperties().remove("io.sniffy.injectHtml");
    sniffyConfiguration.loadSniffyConfiguration();
    sniffyConfiguration.setInjectHtml(false); //outdated
    assertFalse(sniffyConfiguration.isInjectHtml());
}

# Updated by ReAccept
public void testInjectHtml() {
    ... //five behavioral checks omit in the paper due to the page limitation
    // overriden value
    System.getProperties().remove("io.sniffy.injectHtml");
    sniffyConfiguration.loadSniffyConfiguration();
-    sniffyConfiguration.setInjectHtml(false);
    assertFalse(sniffyConfiguration.isInjectHtml());
}

# Updated by MuMuTestUp
public void testInjectHtml() {
    ... //five behavioral checks omit in the paper due to the page limitation
    // overriden value
    System.getProperties().remove("io.sniffy.injectHtml");
    sniffyConfiguration.loadSniffyConfiguration();
-    sniffyConfiguration.setInjectHtml(false);
+    sniffyConfiguration.setInjectHtmlEnabled(false);
    assertFalse(sniffyConfiguration.isInjectHtml());
}
\end{lstlisting}

           
Listing~\ref{fig:motivation_example1} illustrates Limitation~\ding{203} of prior studies.\footnote{From pull request 289 in the Sniffy project: \url{https://github.com/sniffy/sniffy/pull/289}.}
The test method \verb|testInjectHtml()| originally contains six behavioral checks, where the last check invokes \verb|setInjectHtml()| to disable the \texttt{InjectHtml} flag. However, after the pull request is merged, the method \verb|setInjectHtml()| is no longer available, causing the test to fail at compilation. Correctly updating the test requires replacing \verb|setInjectHtml()| with \verb|setInjectHtmlEnabled()|, thereby restoring compilability and executability while preserving coverage of the original code logic.

State-of-the-art approaches such as \reaccept rely on lightweight coverage guidance during test updating. When encountering this compilation error, \reaccept removes the erroneous statement rather than repairing it. Consequently, the updated test retains only the first five behavioral checks and discards the sixth one, resulting in incomplete coverage of the original functionality.
In contrast, \toolname leverages more fine-grained coverage information to correctly update the statement to \texttt{sniffyConfiguration.setInjectHtmlEnabled(false)}. This update not only restores test compilability and executability but also preserves coverage of the original behavior.

We further observe that another SOTA approach \testupdater fails to generate compilable test cases in either motivating example, resulting in inferior performance compared to \reaccept.

%% file: sections/approach.tex
\section{\toolname}


\toolname is a multi-agent framework for automated test case updating. It takes as input the original test case, the focal methods, relevant non-test methods, class and member variables, as well as the diff hunks introduced by a pull request.

The framework updates test cases through four stages. First, \toolname performs input preprocessing to filter redundant and irrelevant contextual information, retaining only context that is relevant to test updating. Second, based on the refined context, \toolname generates an initial updated test case. Third, the updated test is compiled and executed; if execution succeeds, coverage and mutation information is collected. Fourth, \toolname conducts outcome-driven analysis based on the execution results.

Specifically, when test execution fails, the framework analyzes execution logs to identify root causes and generates corresponding repair instructions, retrieving additional contextual information on demand to support accurate fixes. When execution succeeds but coverage falls below predefined thresholds, \toolname analyzes coverage reports and produces instructions to improve line- and branch-level coverage. When the mutation score does not meet the threshold, the framework analyzes mutation reports and generates instructions to strengthen test assertions and improve fault-detection capability.

All instructions produced in the fourth stage are aggregated and fed back to the test update stage, enabling iterative refinement until predefined stopping criteria are satisfied or a maximum number of iterations is reached.

To support this end-to-end workflow, \toolname employs a set of specialized agents responsible for input preprocessing, test updating, error analysis, information retrieval, coverage analysis, and mutation analysis, including the \inputProcessing agent, \testUpdate agent, \errorAnalysis agent, \retrievaler agent, \coverageAnalysis agent, and \mutationAnalysis agent. Each agent is equipped with dedicated tools and prompt templates tailored to its specific task.

\begin{figure}
    \centering
    \includegraphics[width=1\linewidth]{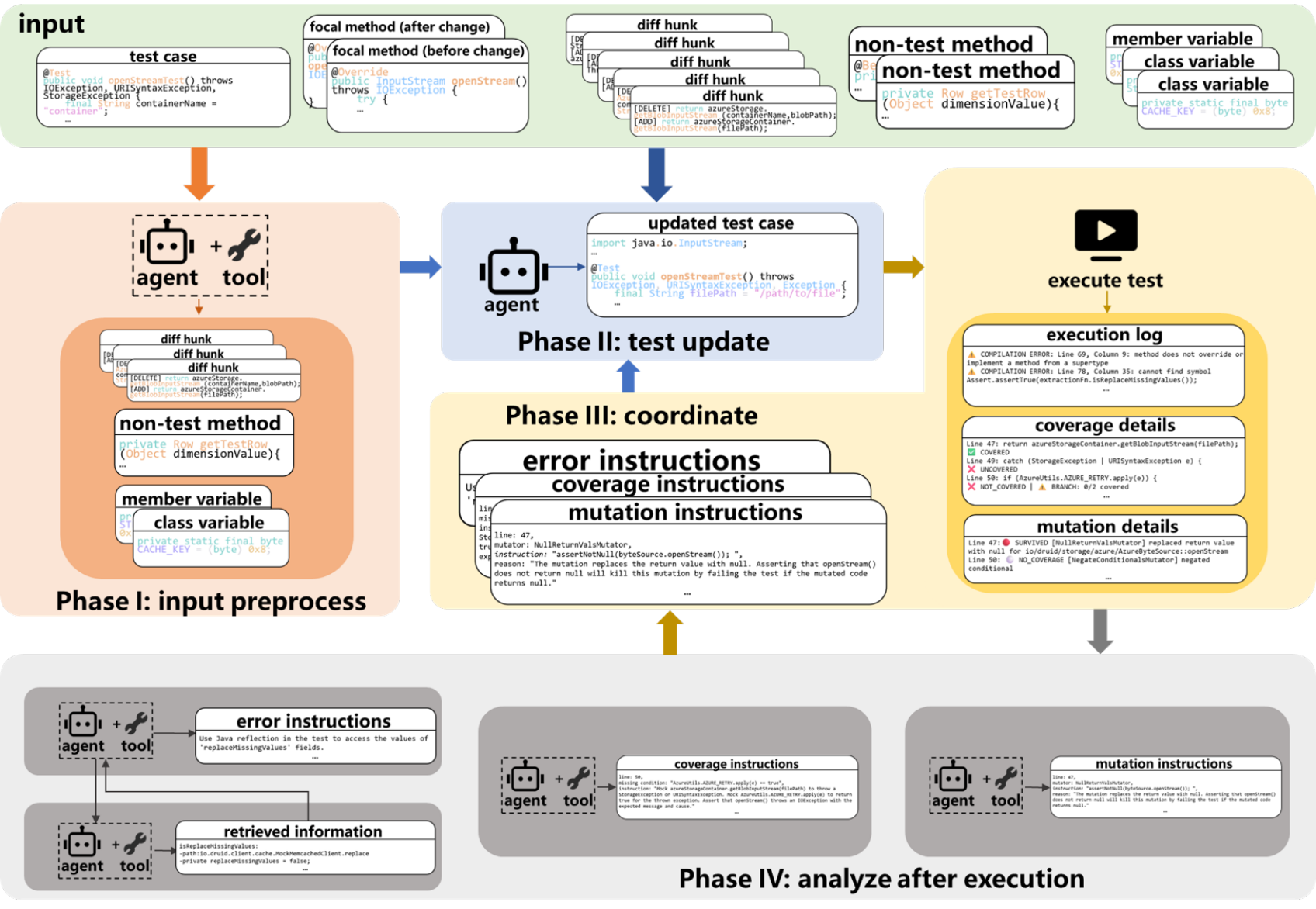}
    \caption{Overview of \toolname.}
    \label{fig:overall_framework}
\end{figure}

Figure~\ref{fig:overall_framework} presents an overview of the framework. In total, \toolname provides 18 tools to support automated test case updating (Table~\ref{tab:tools}), with additional details available in the online appendix. Here, we focus on describing the agents used in \toolname.

\begin{table}[htbp]
  \centering
  \small
  \caption{Overview of tools used in different agents.
  Detailed description is in the replication package.}
    \begin{tabular}{ll}
    \toprule
    Tool used in each agent  & Description \\
    \midrule
    \textbf{\inputProcessing:} & \\
    \midrule
    rank\_hunks & Rank diff hunks related to outdated test code. \\
    \midrule
    \textbf{\coordinator:} & \\
    \midrule
    extract\_test\_code & Extract and write updated test code and import statements. \\
    run\_test\_coverage & Run tests and generate coverage reports. \\
    run\_test\_mutation & Perform mutation testing to assess assertions. \\
    record\_best & Record the best-scoring updated test case. \\
    has\_done & Check whether the update process should terminate. \\
    choose\_agent & Select the appropriate analysis agent. \\
    merge\_instructions & Combine instructions from different agents into key-values format. \\
    \midrule
    \textbf{\errorAnalysis:} & \\
    \midrule
    extract\_error\_info & Extract failure messages from test logs. \\
    distinguish\_unknown\_symbols & Identify unknown symbols in error logs. \\
    locate\_error & Locate the exact line causing an error. \\
    gen\_method\_error\_annotations & Add error annotations to methods. \\
    \midrule
    \textbf{\coverageAnalysis:} & \\
    \midrule
    extract\_uncover\_info & Get uncovered lines and branches from coverage reports. \\
    gen\_method\_coverage\_annotations & Add coverage annotations to methods. \\
    \midrule
    \textbf{\mutationAnalysis:} & \\
    \midrule
    extract\_mutation\_info & Identify surviving mutants from mutation reports. \\
    gen\_method\_mutation\_annotations & Add mutation annotations to methods. \\
    \midrule
    \textbf{\retrievaler:} & \\
    \midrule
    create\_chromaDB & Build ChromaDB for semantic retrieval. \\
    generate\_embedding & Generate vector embeddings for queries. \\
    query\_info & Provide a unified retrieval interface. \\
    \bottomrule
    \end{tabular}%
  \label{tab:tools}%
\end{table}%


\subsection{\inputProcessing} 
The \inputProcessing agent is designed to reduce contextual noise by filtering redundant non-test methods, class and member variables, and diff hunks that are irrelevant to test case updating. This step is necessary because, in practice, only a subset of the available contextual information contributes to effective test updates. Prior studies have shown that, when applying large language models with prompt engineering, excessively verbose prompts often degrade rather than improve performance, and that prompt relevance and quality are more important than prompt size \cite{levy-etal-2024-task,zheng2025an,liu2025effectspromptlengthdomainspecific,leng2024longcontextragperformance}. Motivated by these findings, \toolname explicitly filters contextual information before invoking the test update process.

To identify diff hunks that are relevant to test updating, we draw inspiration from the work of Yaraghi et al.~\cite{TARGET}, who rank diff hunks using multiple signals, including TF-IDF similarity between manually updated test snippets and diff hunks (Breakage TF-IDF Similarity), diff hunk frequency (Repetition), and TF-IDF similarity between the full outdated test case and diff hunks (Test TF-IDF Similarity). However, Breakage TF-IDF Similarity requires access to manually updated tests to identify update locations, which is unrealistic in a fully automated setting because it implicitly exposes the model to partial ground truth. To avoid this issue, we rely only on Repetition and Test TF-IDF Similarity, and use the \verb|rank_hunks| tool to rank and select diff hunks without assuming access to manual updates.

In addition to diff hunks, non-test methods and class or member variables may also be required for correctly updating and executing test cases. While prior work largely overlooks this information, we explicitly incorporate it into the preprocessing stage. Specifically, we first collect all non-test methods and class/member variables defined in the test class. Since not all of them are relevant to the test update, we further filter this information by instructing the LLM to retain only the necessary elements. Concretely, we construct a prompt that includes the original test case, the focal methods, candidate non-test methods, class/member variables, and the filtered diff hunks, and ask the LLM to discard irrelevant non-test methods and variables (the prompt is in the replication package).

finally, the \inputProcessing agent produces a refined set of non-test methods, class/member variables, and diff hunks that provide concise and relevant context for test case updating.

\subsection{\testUpdate} 
Based on the filtered context, the \testUpdate agent performs the actual test case update. It takes as input the outdated test case, the focal methods, the contextual information produced by the \inputProcessing agent, and the update instructions generated by the \coordinator agent, which is described in Section~\ref{sec:coordinator}. During the initial update, when no instructions from the \coordinator agent are available, the LLM is instructed to perform minimal modifications to restore correctness. To ensure that the updated test can be compiled and executed, the agent is also instructed to generate any required import statements. The output of this stage is an updated test case together with the corresponding import declarations.

\subsection{\coordinator} \label{sec:coordinator}

The \coordinator agent orchestrates the execution of updated test cases and determines subsequent analysis steps based on the observed execution outcomes. It serves as the central control component that connects test updating with feedback-driven refinement.

After the \testUpdate agent produces an updated test case, the output may contain redundant or auxiliary information due to the generative nature of large language models. To ensure that only valid test code is applied, the \coordinator agent first invokes the \verb|extract_test_code| tool to extract the updated test case and the corresponding import statements, and then applies them to the code repository.

To assess whether the updated test preserves sufficient testing functionality, the agent executes the test and collects quality signals for the focal method. Specifically, it invokes \verb|run_test_coverage| and \verb|run_test_mutation| to obtain line- and branch-level coverage information as well as mutation scores. These signals provide the perspectives on test adequacy beyond mere executability.

Because the test updating process is iterative and influenced by the inherent randomness of LLMs, later iterations may inadvertently degrade test quality. To mitigate this issue, the \coordinator agent invokes the \verb|record_best| tool to maintain the historically best-performing test case observed so far. The agent then checks termination conditions by invoking \verb|has_done|, which determines whether the current test satisfies the predefined coverage and mutation thresholds, or whether the maximum number of iterations has been reached.

When further refinement is required, the \coordinator agent selects appropriate analysis agents based on the execution results. For instance, if compilation errors or assertion failures occur, it triggers the \errorAnalysis agent; if line or branch coverage falls below the threshold, it invokes the \coverageAnalysis agent; and if the mutation score is insufficient, it activates the \mutationAnalysis agent. Each selected agent analyzes a specific aspect of test inadequacy and produces corresponding feedback.

After all selected analysis agents finish, the \coordinator agent aggregates their outputs into a unified set of update instructions. This aggregation is performed using the \verb|merge_instructions| tool, which consolidates feedback from different agents into a structured key–value format, including error analysis results, coverage analysis results, and mutation analysis results. The merged instructions are then fed back to the test updating stage to guide the next iteration.

\subsection{\errorAnalysis} 

The \errorAnalysis agent analyzes execution error logs produced during test execution and generates corresponding repair instructions. In our setting, test cases are compiled and executed using Maven, whose execution logs often contain substantial redundant information that is unrelated to the root causes of failures. To reduce noise and focus the analysis, the agent first invokes the \verb|extract_error_info| tool to extract concise error messages from the raw logs, and then distinguishes between compilation errors and assertion failures.

When a compilation error is detected, the agent parses the error messages to identify the symbols involved in the failure. It then invokes \verb|distinguish_unknown_symbols| to classify these symbols. This matching-based tool determines whether a symbol belongs to a predefined set of known symbols, such as those from standard libraries or testing frameworks. Symbols not appearing in this set are classified as unknown.

For known symbols, the agent generates repair instructions that add the corresponding import statements. For unknown symbols, which may either correspond to project-specific code or result from hallucinations introduced by the LLM, the agent searches the code repository with the assistance of the \retrievaler agent. If the symbol is found, the agent generates instructions to add the appropriate import. If only semantically similar symbols are retrieved, the agent corrects the hallucinated symbol and produces the repair instructions, including the necessary imports.

When the failure is caused by an assertion error, the agent extracts the assertion failure message together with the expected and actual values reported in the logs. Based on this information, the LLM generates repair instructions that revise the assertion to reflect the updated behavior.

To further improve the accuracy of the generated instructions, the \errorAnalysis agent explicitly localizes error positions in the test code. In both compilation and assertion failure scenarios, the agent invokes \verb|locate_error| to identify the precise error location from the execution logs, and then uses \verb|gen_method_error_annotations| to insert error-related comments directly into the test code. By making the LLM aware of where the error occurs, the agent helps the model focus on the relevant context and avoid unnecessary modifications. For example, in Figure~\ref{fig:motivation_example2}, the outdated test fails to compile at the second-to-last line, and a comment such as “cannot find method: \texttt{setInjectHtml()}” is inserted at that location. Assertion failures are handled in a similar manner by annotating the corresponding assertion statements.

Finally, based on the analyzed error type and localized context, the \errorAnalysis agent outputs structured repair instructions that guide subsequent test updates.

\subsection{\coverageAnalysis} 

The \coverageAnalysis agent analyzes coverage reports and generates guidance for exercising uncovered code in the focal method.

Although coverage reports contain extensive information, the agent focuses exclusively on lines and branches within the focal method that remain uncovered. To identify these elements, it first invokes the \verb|extract_uncover_info| tool to extract uncovered lines and branches from the coverage report.
To help the LLM reason about the uncovered code in context, the agent then invokes \verb|gen_method_coverage_annotations| to produce an annotated version of the focal method. Each line is labeled as \texttt{COVERED}, \texttt{NOT\_COVERED}, or \texttt{NO\_INSTRUCTION}, indicating covered lines, uncovered lines, and non-executable lines, respectively. Branch annotations further capture partial coverage information in the form \texttt{BRANCH: i/n covered}, where n denotes the total number of branches in a statement and i indicates the number of branches that have been executed. These annotations allow the LLM to directly observe which execution paths remain unexplored.

Recognizing that different uncovered lines and branches vary in difficulty to exercise, the agent prompts the LLM to distinguish between those that are relatively easy to cover and those that are more challenging. Based on this classification and the annotated context, the \coverageAnalysis agent produces coverage-improvement instructions for uncovered lines and branches, guiding the test update process toward more comprehensive functional coverage.

\subsection{\mutationAnalysis} 

The \mutationAnalysis agent complements coverage analysis by examining mutation reports and generating guidance for strengthening test assertions.

While mutation reports contain numerous details, the agent concentrates on mutants associated with the focal method that remain alive. It first invokes the \verb|extract_mutation_info| tool to identify surviving mutants, and then uses \verb|gen_method_mutation_annotations| to annotate the focal method accordingly.
In the annotated code, each line containing mutants is labeled with the status of each mutant, including \texttt{NO\_COVERAGE} (the line is not covered), \texttt{SURVIVED} (the mutant remains alive), or \texttt{KILLED} (the mutant has been eliminated). Each annotation also records the mutation operator and a brief description, and is inserted as a comment adjacent to the corresponding line. These annotations make mutation outcomes explicit and localized for the LLM.
Based on this information, the agent distinguishes between uncovered mutants, which require improved functional coverage, and surviving mutants, which indicate insufficient assertion strength. The \mutationAnalysis agent then generates instructions for killing surviving mutants by strengthening or introducing assertions, thereby guiding the test case update process toward improved oracle robustness and fault-detection capability.

\subsection{\retrievaler} 

The \retrievaler agent supports error analysis by resolving unknown symbols encountered during test updating and providing relevant contextual information to the \errorAnalysis agent. When a symbol is successfully retrieved, the agent returns the corresponding context: for methods, this includes the method signature and file path; for fields, it includes the field definition and its location. This retrieved context enables the \errorAnalysis agent to correctly reference, invoke, and import the resolved symbols during test repair.

To enable efficient semantic retrieval, the agent first invokes \verb|create_chromaDB| to construct a ChromaDB vector database over the project code. To control the embedding cost, the agent does not embed the entire project. Instead, it relies on the LLM to select a single relevant module for embedding. Within the selected module, the agent embeds natural language descriptions of methods and fields, such as Javadoc comments, which provide concise semantic representations suitable for retrieval.

Given an unknown symbol, the LLM generates a query text describing the symbol’s intended semantics. The agent then invokes \verb|generate_embedding| to embed the query and uses \verb|query_info| to compute cosine similarity between the query embedding and entries in the vector database. Retrieved candidates are subsequently evaluated by the LLM, which filters out irrelevant results and retains only context deemed useful for symbol resolution.

In cases where a symbol cannot be resolved in a single retrieval step, the agent performs additional retrieval iterations. During each iteration, the LLM decides whether to expand the search by embedding an additional module or to refine the query text. This iterative strategy accounts for two common failure modes: the symbol may reside outside the currently embedded module, or the initial query text may provide an incomplete or biased semantic description. The process continues until all symbols are resolved or a predefined maximum number of retrieval iterations is reached. In practice, our experiments show that even with multiple retrieval rounds, the overall cost of test case updating remains manageable.

Ultimately, the \retrievaler agent returns the retrieved contextual information with a list of symbols that could not be resolved, allowing downstream agents to handle them explicitly.

%% file: sections/data.tex
\section{Dataset}

To evaluate the effectiveness of \toolname in the real-world scenario, we construct a dataset at the pull-request level.
This is because prior studies, e.g., \tarbench, focused on single-commit scenarios, overlooking the fact that test updates are often influenced by changes introduced across multiple commits.
In practice, updates to test code to accommodate evolving production functionality often span multiple commits, and commits within a single pull request typically correspond to a coherent software engineering task, such as implementing a new feature or fixing a bug.
These observations motivate us to build a new, comprehensive, and high-quality pull-request–level dataset.

To construct the dataset, we start from the 59 Java projects provided by \tarbench, which have been validated to satisfy the quality requirements for test case updating and therefore form a reliable initial project set. For each project, we use the GitHub API to retrieve all merged pull requests. We treat the commit immediately preceding the first commit in a pull request as the pre-PR version, and the last commit in the pull request as the post-PR version. Using git diff, we identify pull requests that involve changes to test code.

Next, we identify the focal methods of test cases, i.e., the production methods targeted by the tests. Identifying focal methods allows us to restrict coverage and mutation analysis to the relevant functionality rather than the entire codebase, thereby providing more precise and actionable feedback for guiding test case updates. In contrast, measuring coverage or mutation scores at the repository or class level may introduce substantial noise from unrelated code and mislead the update process.

Most prior datasets \cite{SITAR,CEPROT,REACCEPT,CommitUp,TestUpdater} identify focal methods but restrict their scope to cases where the focal method itself is modified. Such a design overlooks a broader class of test-outdated scenarios, for example, when changes occur in the call chain while the focal method remains unchanged. To capture these scenarios, we do not require the focal method to be directly modified.

To identify the focal methods of the tests, for each pull request, we identify focal methods by combining call-graph analysis and method name matching. Following prior studies \cite{TestUpdater,CommitUp,REACCEPT}, a focal method is defined as a production method that (1) is directly invoked by the test case, (2) shares the same directory structure with the test case except for the test and main segments, and (3) exhibits the highest method name similarity.
For example, the test method
\url{wdtk-rdf/src/test/java/org/wikidata/wdtk/rdf/PropertyRegisterTest.java#testSetMissingPropertyTypeFromStringValue()}
is matched to the focal method
\url{wdtk-rdf/src/main/java/org/wikidata/wdtk/rdf/PropertyRegister.java#setPropertyTypeFromStringValue()}.

To identify outdated test cases, we follow the procedure proposed in \cite{TARGET} and perform three rounds of compilation and execution: (1) Execute the pre-PR tests on the pre-PR version; (2) Execute the pre-PR tests on the post-PR version; and (3) Execute the post-PR tests on the post-PR version.
During the first two executions, we collect coverage and mutation information for the focal methods. Unlike prior work, which primarily considers compilation errors and test failures, we additionally account for cases where test cases become outdated due to degraded focal-method coverage or mutation scores. Specifically, a test case is considered outdated if (1) the first execution succeeds, (2) the second execution either fails to compile, fails at runtime, or exhibits reduced focal-method coverage or mutation score, and (3) the third execution succeeds.

Applying this procedure yields a dataset of 571 samples from 10 open-source Java projects. Table~\ref{tab:dataset} summarizes the key statistics of \dataset, together with the causes of test obsolescence identified in the second execution.
Although outdated test cases due to coverage or mutation degradation are relatively rare compared to those caused by compilation errors or test failures, considering these signals remains important for preserving the functional adequacy and robustness of updated test cases.

\begin{table}[htbp]
  \centering
  \caption{Statistics of \dataset. The table summarizes the outdated test cases identified during the second execution, categorized by cause. (LOC: lines of code; Class: average class; TC: average test class; PR: pull request; Error: compile error; Failure: test failure; CD: coverage degradation; MD: mutation score degradation).}
    \begin{tabular}{cccccccccc}
    \toprule
    projects & LOC & \# Class & \# TC & \# PR & \# Error & \# Failure & \# CD & \# MD & \# Sample\\
    \midrule
    Wikidata & 50623 & 476	 & 631  & 30    & 83    & 23    & 1     & 0     & 107 \\
    druid & 244072	 & 1989	 & 1493  & 23    & 30    & 3     & 0     & 0     & 33 \\
    cucumber & 5720	 & 723	 & 167  & 46    & 48    & 67    & 1     & 1     & 117 \\
    hutool & 255370 & 2075	 & 2464  & 6     & 11    & 1     & 2     & 3     & 17 \\
    ormlite & 48767	 & 946	 & 1046  & 2     & 1     & 0     & 2     & 0     & 3 \\
    mockserver & 183807	 & 1122	 & 3532  & 1     & 0     & 1     & 0     & 0     & 1 \\
    neo4j & 49726 & 444	 & 784  & 47    & 80    & 28    & 1     & 0     & 109 \\
    pac4j & 42465 & 823	 & 695  & 19    & 141   & 12    & 1     & 2     & 156 \\
    sniffy & 17098 & 201	 & 4678  & 7     & 7     & 18    & 1     & 0     & 26 \\
    nfe   & 164532	 & 1933	 & 3575  & 2     & 0     & 2     & 0     & 0     & 2 \\
    \midrule
    total &   -    &   -    &    -   & 183   & 401   & 155   & 9     & 6     & 571 \\
    \bottomrule
    \end{tabular}%
  \label{tab:dataset}%
\end{table}%

%% file: sections/experiment.tex
\section{Experimental Setup}





\subsection{Evaluation Metrics}\label{sec:metrics}

Following prior work on test case updating~\cite{REACCEPT,TestUpdater} and test case generation~\cite{Panichella_dynamosa_tse2018}, we evaluate test behavior and effectiveness using five metrics: Compilation Pass Rate (CPR), Test Pass Rate (TPR), Line Coverage (Line Cov), Branch Coverage (Branch Cov), and Mutation Score (Mut Score).
CPR measures the proportion of updated tests that compile successfully, reflecting syntactic correctness. TPR captures the proportion of updated tests that pass on the updated code, indicating functional correctness. Line Cov and Branch Cov measure the extent to which executable lines and control-flow branches in the focal methods are exercised, respectively. Mut Score measures the proportion of injected mutants killed by the updated tests, reflecting assertion strength and fault-detection capability.



\subsection{Baselines and Configurations}

We compare \toolname with three state-of-the-art baselines for automated test case updating. \target~\cite{TARGET} formulates test updating as a code translation task and applies a pretrained code language model (CodeT5+). \reaccept~\cite{REACCEPT} is an LLM-based approach that integrates iterative execution feedback with retrieval-augmented generation to restore test executability and cover newly introduced code. \testupdater~\cite{TestUpdater} further incorporates language-server–based post-hoc retrieval and explicitly considers coverage of both changed and unchanged code, achieving strong compilation and execution performance.

For \target, we adopt the best-performing open-source pretrained model recommended by the authors.\footnote{\url{https://figshare.com/articles/dataset/_b_TaRBench_A_Comprehensive_Benchmark_for_Automated_Test_Case_Repair_b_/25008893}}
As \target supports only single-hunk updates, we concatenate multiple hunks into a single hunk when necessary. We follow the remaining configurations suggested by \target, except that we disable Breakage TF-IDF Similarity for hunk ranking, as it relies on partial manually updated information and would introduce unfair advantages.

We evaluate both closed-source and open-source LLMs, specifically GPT-4.1 and DeepSeek-V3.2. For all models, the temperature is set to 0 to ensure reproducibility. Following prior work~\cite{TestUpdater}, we limit test updating to a maximum of four iterations. Coverage is measured using JaCoCo, and mutation analysis is performed using PIT with all standard mutation operators enabled. Unless otherwise stated, the thresholds for line coverage, branch coverage, and mutation score are set to 100\%, and the maximum number of retrieval iterations is set to three.


%% file: sections/result.tex
\section{Experimental Results}

We evaluate the effectiveness of our framework \toolname with the following research questions:

\begin{itemize}[wide=0pt]
    \item \textbf{RQ1: Compilation and Test Pass Rate.} How does \toolname perform in terms of Compilation Pass Rate (CPR) and Test Pass Rate (TPR)?
    \item \textbf{RQ2: Coverage and Mutation.} How does \toolname perform in terms of Line Coverage (Line Cov), Branch Coverage (Branch Cov), and Mutation Score (Mut Score)?
    \item \textbf{RQ3: Ablation Study.} What are the contributions of \toolname’s core components?
\end{itemize}

\subsection{RQ1: Compilation and Test Pass Rate}



\begin{table}[htbp]
\small
  \centering
  \caption{Comparison between our approach and all baselines in terms of Compilation Pass Rate (CPR) and Test Pass Rate (TPR).}
    \begin{tabular}{cccc}
    \toprule
    \textbf{Model} & \textbf{Method} & \textbf{CPR (\%)} & \textbf{TPR (\%)} \\
    \midrule
    CodeT5+ & \target & 56.39 & 46.41 \\
    \midrule
    \multirow{3}[2]{*}{GPT-4.1} & \reaccept & 75.13 & 62.17 \\
          & \testupdater & 66.02 & 56.22 \\
          & \toolname  & \textbf{84.24} & \textbf{78.28} \\
    \midrule
    \multirow{3}[2]{*}{Deepseek-v3.2} & \reaccept & 78.81 & 63.75 \\
          & \testupdater & 67.25 & 59.54 \\
          & \toolname  & \textbf{84.06} & \textbf{80.91} \\
    \bottomrule
    \end{tabular}%
  \label{tab:rq1}%
\end{table}%

Table~\ref{tab:rq1} reports the comparison of Compilation Pass Rate (CPR) and Test Pass Rate (TPR) between \toolname and three baseline approaches under different backend models. Overall, \toolname consistently achieves the best performance across all evaluated settings.

Compared with \target, \toolname shows substantial improvements under both GPT-4.1 and DeepSeek-V3.2. With GPT-4.1, \toolname improves CPR by 49.39\% and TPR by 68.67\%, while with DeepSeek-V3.2 it improves CPR by 49.07\% and TPR by 74.34\%. These gains can be largely attributed to the ability of \toolname to handle both single-hunk and multi-hunk updates, which frequently arise in real-world development. In contrast, \target is specifically designed for single-hunk updates and therefore struggles when multiple hunks need to be updated jointly.

\toolname also consistently outperforms \reaccept and \testupdater. When using GPT-4.1, \toolname improves over \reaccept by 12.13\% in CPR and 25.91\% in TPR. This improvement is primarily due to \toolname’s ability to retrieve and incorporate project-specific contextual information required for correct test updates, ensuring that updated tests can ultimately compile and execute successfully. In contrast, \reaccept relies on retrieval-augmented generation mainly to obtain historical samples for few-shot prompting, but does not retrieve concrete contextual information from the target codebase.

Under GPT-4.1, \toolname further outperforms \testupdater by 27.60\% in CPR and 39.24\% in TPR. This advantage stems from \toolname’s more robust retrieval mechanism, which resolves hallucinated symbols through semantic search. By contrast, although \testupdater retrieves contextual information, it relies solely on exact matching, which limits its ability to handle hallucinated symbols generated by LLMs and may lead to incorrect retrieval and misguided updates.

Similar trends are observed when using DeepSeek-V3.2. In this setting, \toolname again outperforms \reaccept by 6.66\% in CPR and 26.92\% in TPR, and outperforms \testupdater by 25.00\% in CPR and 35.89\% in TPR. These results indicate that \toolname consistently outperforms prior approaches across both closed-source and open-source LLMs, demonstrating robustness with respect to the choice of backend model.

In addition to its effectiveness, \toolname incurs moderate overhead in both runtime and cost. The average runtime per test case is 41.56s with GPT-4.1 and 130.70s with DeepSeek-V3.2, which is approximately 2$\times$ that of the baselines (17.39s/20.72s for \reaccept and 70.23s/82.59s for \testupdater). Notably, mutation analysis (PITest) and coverage analysis (JaCoCo) contribute only a small portion of the total runtime (e.g., 14.39\%/4.26\% with GPT-4.1 and 7.66\%/2.04\% with DeepSeek-V3.2), indicating that the overhead is primarily introduced by multi-agent coordination rather than auxiliary analyses.

In terms of cost, the average cost of updating a single test case is \$0.05952 with GPT-4.1 and \$0.0064 with DeepSeek-V3.2, which is about 3$\times$ higher than the baselines (\$0.02129/\$0.00215 for \reaccept and \$0.01384/\$0.00173 for \testupdater). This increase is expected due to the additional interactions required by multi-agent coordination. However, the overall cost remains relatively low in absolute terms and falls within a practically acceptable range. Moreover, compared to prior multi-agent approaches such as SWE-agent~\cite{yang2024swe}, which report 8–13$\times$ cost increases over single-agent baselines, \toolname demonstrates a more favorable cost–performance trade-off.

Overall, \toolname substantially improves the quality of updated test cases while maintaining a controlled and practical balance between effectiveness and efficiency.

\begin{tcolorbox}[RQBox, title={Answer RQ1}]
\toolname achieves substantial improvements in both Compilation Pass Rate and Test Pass Rate regardless of whether GPT-4.1 or DeepSeek-V3.2 is used.
\end{tcolorbox}

\subsection{RQ2: Coverage and Mutation}


\begin{table}[htbp]
  \centering
  \small
  \caption{Comparison between our approach and all baselines in terms of Line Coverage (Line Cov), Branch Coverage (Branch Cov), and Mutation Score (Mut Score).}
    \begin{tabular}{ccccc}
    \toprule
    \textbf{Model} & \textbf{Method} & \textbf{Line Cov (\%)} & \textbf{Branch Cov (\%)} & \textbf{Mut Score (\%)} \\
    \midrule
    CodeT5+ & \target & 81.4  & 43.29  & 56.35 \\
    \midrule
    \multirow{3}[2]{*}{GPT-4.1} & \reaccept & 82.81 & 48.28  & 56.71 \\
          & \testupdater & 84.44 & 52.83  & 62.05 \\
          & \toolname  & \textbf{88.94} & \textbf{63.36} & \textbf{72.39} \\
    \midrule
    \multirow{3}[1]{*}{Deepseek-v3.2} & \reaccept & 84.09 & 49.15  & 57.87 \\
          & \testupdater & 86.21 & 51.38  & 62.81 \\
          & \toolname  & \textbf{88.66} & \textbf{63.72} & \textbf{74.33} \\
    \bottomrule
    \end{tabular}%
  \label{tab:rq2}%
\end{table}%

Table~\ref{tab:rq2} reports the line coverage, branch coverage, and mutation score achieved by all compared approaches. Overall, \toolname consistently attains the best results across all three metrics. When using GPT-4.1, \toolname outperforms the strongest baseline, \testupdater, by 5.33\% in line coverage, 19.93\% in branch coverage, and 16.66\% in mutation score. These gains indicate that explicitly incorporating branch-coverage guidance enables \toolname to more effectively exercise branch logic, which is largely ignored by existing approaches.

Moreover, \toolname yields a notable improvement in mutation score (16.66\% over the best baseline). This result suggests that explicitly incorporating mutation-guided feedback helps preserve and enhance assertion robustness in updated tests, which is not directly addressed by existing approaches.

Similar trends are observed under DeepSeek-V3.2. In this setting, \toolname again outperforms \testupdater by 2.84\% in line coverage, 24.02\% in branch coverage, and 18.34\% in mutation score. Taken together, these results demonstrate that \toolname substantially improves both coverage and mutation effectiveness, regardless of the underlying LLM.




\begin{tcolorbox}[RQBox, title={Answer RQ2}]
\toolname improves both coverage and mutation effectiveness through joint coverage- and mutation-guided test updating.
\end{tcolorbox}

\subsection{RQ3: Ablation Study}\label{sec:ablation}

Here, we use GPT-4.1 as a backend LLM to conduct an ablation study, as it and Deepseek-V3.2 have comparable performance.



We design four ablation settings to evaluate the contribution of key components in addressing the limitations discussed in Section~\ref{sec:intro}. Specifically, to assess Limitation~\ding{202}, which concerns the adequacy of test assertions, we disable mutation-guided assertion updating by removing the entire \mutationAnalysis agent. To evaluate Limitation~\ding{203}, which highlights the lack of branch-level guidance in prior work, we restrict coverage guidance to line coverage only by modifying the \coverageAnalysis agent to exclude branch coverage analysis and branch-level repair instructions. To assess Limitation~\ding{204}, which relates to reliance on exact matching for context retrieval, we replace semantic similarity–based retrieval with language-server–based exact matching by modifying the \retrievaler agent to adopt the retrieval strategy used in \testupdater. 
Finally, to quantify the overall contribution of our approach beyond the use of a powerful LLM, we introduce a fourth ablation setting using a vanilla GPT-4.1 model without any of our proposed agents or guidance mechanisms. This setting allows us to explicitly measure how much performance gain is attributable to our framework design rather than the raw capability of the underlying LLM.


\begin{table}[htbp]
  \centering
  \small
  \caption{Results of ablation study on GPT-4.1. ``w/o mut" indicates ``removing the entire\mutationAnalysis agent", ``w/o branch" indicates ``the \coverageAnalysis agent disables branch coverage analysis and guidance", ``w language server" indicates ``the \retrievaler agent adopt language-server–based exact-match retrieval strategy". The best result is highlighted in bold.}
    \begin{tabular}{cccccc}
    \toprule
    \textbf{Method} & \textbf{CPR (\%)} & \textbf{TPR (\%)} & \textbf{Line Cov (\%)} & \textbf{Branch Cov (\%)} & \textbf{Mut Score (\%)} \\
    \midrule
    \toolname  & \textbf{84.24} & \textbf{78.28} & 88.94 & 63.36 & \textbf{72.39} \\
    w/o mut & 81.61 & 76.71 & 89.54 & \textbf{66.87} & 65.88 \\
    w/o branch & 82.31 & 78.46 & \textbf{89.60} & 61.11 & 72.31 \\
    w language server & 79.51 & 74.08 & 87.77 & 65.37  & 71.45 \\
    vanilla GPT-4.1 & 61.65 & 48.86 & 86.94 & 51.40 & 61.20 \\
    \bottomrule
    \end{tabular}%
  \label{tab:rq3}%
\end{table}%

\paragraph{\toolname v.s. w/o mut: }

When disabling the \mutationAnalysis agent, the mutation score of \toolname decreases by 8.99\%, indicating that \mutationAnalysis agent plays a critical role in strengthening the mutation-killing capability of updated tests and improving assertion robustness. In addition, the absence of \mutationAnalysis results in a 3.12\% reduction in CPR and a 2.01\% reduction in TPR. This suggests that, without explicit assertion guidance, LLMs tend to generate or modify assertions in ways that increase the likelihood of producing invalid or inconsistent assertions and, consequently, of compilation failures or assertion violations.

Interestingly, disabling \mutationAnalysis yields slight improvements in line coverage (0.67\%) and branch coverage (5.54\%). While the gain in line coverage is modest, the improvement in branch coverage is non-negligible. This phenomenon can be explained by the interaction between coverage- and mutation-guided feedback during iterative updating. When \mutationAnalysis is enabled, many test cases simultaneously fail to satisfy both coverage and mutation thresholds, causing \toolname to receive guidance from both analyses and thus divide the model’s attention. In contrast, without mutation analysis, the update process receives only coverage-oriented instructions, allowing the LLM to focus exclusively on improving coverage and thereby achieving higher branch coverage.
Nevertheless, higher coverage without sufficiently strong assertions provides limited practical value, as it does not guarantee correct behavioral validation. Consequently, a modest reduction in branch coverage in exchange for a substantial improvement in mutation score represents a more desirable trade-off in practice.

\paragraph{\toolname v.s. w/o branch: }

When branch coverage analysis and guidance are disabled in the \coverageAnalysis agent, the branch coverage of \toolname decreases by 3.55\%, indicating that explicit branch-oriented guidance plays an important role in enabling updated test cases to exercise complex control-flow paths and corner-case behaviors. In addition, CPR decreases by 2.29\%, suggesting that branch-coverage guidance helps LLMs explore conditional logic that would otherwise remain untested.

Disabling branch coverage guidance also results in slight increases in TPR and line coverage. The marginal improvement in TPR can be attributed to the fact that exercising complex branch logic often requires generating more sophisticated assertions, which are more challenging for LLMs to synthesize correctly and may introduce assertion errors. However, this increase in TPR does not necessarily reflect improved test quality: the mutation score decreases, indicating weakened assertion strength due to insufficient validation of complex behaviors. The modest increase in line coverage can be explained by a shift in focus, as removing branch-level guidance allows the model to concentrate on covering uncovered lines rather than balancing both line- and branch-level objectives.
Overall, the degradation in branch coverage and CPR outweighs the marginal gains in line coverage and TPR, underscoring the importance of incorporating branch coverage analysis and guidance for effective test case updating.

\paragraph{\toolname v.s. w language server: }

When semantic similarity–based retrieval is replaced with language-server–based exact matching, the performance of \toolname decreases by 5.61\% in CPR, 5.37\% in TPR, 1.32\% in line coverage, and 1.30\% in mutation score. The degradation in CPR and TPR suggests that semantic retrieval is more effective at resolving relevant contextual information that cannot be reliably recovered through exact matching, including private symbols and semantically plausible but non-existent (hallucinated) symbols. By retrieving such information, semantic retrieval provides a more complete context to the LLM, thereby improving compilation success and test pass rates.

Although branch coverage increases by 3.17\% when language-server–based retrieval is used, the consistent performance degradation observed across the remaining metrics indicates that semantic similarity–based retrieval offers a more robust and practically effective solution.



\paragraph{\toolname v.s. vanilla GPT-4.1: } 

The vanilla GPT-4.1 baseline results in a 26.82\% decrease in CPR, a 37.58\% decrease in TPR, a 2.25\% decrease in line coverage, an 18.88\% decrease in branch coverage, and a 15.46\% decrease in mutation score. This consistent degradation across all metrics demonstrates that our approach achieves substantial gains beyond direct LLM usage. Furthermore, these results suggest that the improvements stem from our method design rather than potential data leakage.

\begin{tcolorbox}[RQBox, title={Answer RQ3}]
Fine-grained mutation analysis, fine-grained coverage analysis, and two-stage retrieval jointly contribute to the effectiveness of \toolname.
\end{tcolorbox}

%% file: sections/discussion.tex
\section{Discussion}

\subsection{Coverage and Mutation of Jointly Passed Tests}

Coverage and mutation metrics can be computed over different subsets of test cases across approaches, as these metrics require updated tests to compile and execute successfully. To enable a fair comparison, we additionally evaluate line coverage, branch coverage, and mutation score on jointly passed test cases, that is, test cases that are successfully executed by both \toolname and the corresponding baseline.

Due to space constraints, we omit detailed results for all jointly passed settings. Nevertheless, the observed trends are consistent with those reported in the main experiments. Across all evaluated configurations, \toolname consistently achieves the highest line coverage, branch coverage, and mutation score, outperforming the strongest baseline by 5.05\%, 25.13\%, and 25.60\%, respectively.

The ablation study exhibits similar patterns under the jointly passed setting. In particular, disabling mutation analysis leads to an 8.17\% decrease in mutation score relative to \toolname, while disabling branch-level coverage guidance results in a 5.35\% reduction in branch coverage. Replacing semantic retrieval with language-server–based retrieval yields a 5.04\% increase in branch coverage; however, this isolated gain does not compensate for the substantial degradation in compilation success and test pass rate observed in Section~\ref{sec:ablation}, both of which are prerequisites for effective test case updating.

Overall, the jointly passed results further confirm the robustness of \toolname, demonstrating that its advantages are not an artifact of differences in executable sample sets but stem from its stronger ability to improve both functional coverage and assertion robustness.

\subsection{Data Leakage}

\begin{table}[htbp]
  \centering
  \caption{Comparison of n-gram overlap (n=4) across methods for data leakage analysis. Lower is better.}
    \begin{tabular}{ccc}
    \toprule
    Method & Model & n-gram \\
    \midrule
    \multirow{2}[1]{*}{\toolname} & GPT-4.1 & 0.34 \\
          & DeepSeek-V3.2 & 0.31 \\
    \multirow{2}[0]{*}{\reaccept} & GPT-4.1 & 0.54 \\
          & DeepSeek-V3.2 & 0.61 \\
    \multirow{2}[1]{*}{\testupdater} & GPT-4.1 & 0.67 \\
          & DeepSeek-V3.2 & 0.69 \\
    \bottomrule
    \end{tabular}%
  \label{tab:data_leakage}%
\end{table}%

To assess the potential risk of data leakage when using LLMs, we conduct an n-gram overlap analysis following \cite{fan2025exploring}, where $n=4$. This metric measures the lexical overlap between generated outputs and the training corpus, with higher values indicating a greater likelihood of memorization.

As shown in Table~\ref{tab:data_leakage}, \toolname consistently exhibits lower n-gram overlap than the baselines across both GPT-4.1 and DeepSeek-V3.2. For example, under GPT-4.1, \toolname achieves an overlap of 0.34, compared to 0.54 for \reaccept and 0.67 for \testupdater. A similar trend is observed with DeepSeek-V3.2.

These results indicate that \toolname has a lower risk of data leakage. Moreover, if the observed performance gains were primarily driven by memorization, a vanilla GPT-4.1 model would be expected to achieve comparable results. However, this is not the case (see Section~\ref{sec:ablation}), further suggesting that the improvements stem from our method design rather than data leakage.





\subsection{Usage and Scope of \toolname}

\toolname is designed to be applied after production code changes but before submitting a PR. It can be triggered when outdated tests are detected, either manually or via automated tools (e.g., \cite{REACCEPT}). In practice, test failures handled by \toolname may stem from either outdated tests or incorrect production code.

To mitigate the risk of propagating incorrect updates, \toolname operates within the standard PR workflow, where all changes are subject to developer review. As a result, incorrect test or code updates can be identified and filtered during the review process before merging. Consequently, \toolname does not inherently introduce erroneous updates into the codebase. Instead, it facilitates maintaining test–code consistency and reduces the likelihood of PR rejection caused by outdated tests.

\subsection{Threats to Validity}

\paragraph{Internal Validity.}
The dataset used in our experiments, \dataset, is collected from publicly available code repositories. Although \dataset is constructed to reflect realistic pull-request-level software evolution and covers a wide range of test-update scenarios, it may still not fully capture all possible changes encountered in practice. 
Moreover, compared with most prior studies that evaluate on only 200–300 samples, the \dataset contains 571 test update instances, enabling a broader coverage of diverse and complex test update scenarios and reducing evaluation bias.

\paragraph{External Validity.}
The generalizability of our approach may be limited by the target ecosystem. \toolname is currently evaluated on projects using the Maven framework and Java language; these projects may not fully represent software systems built with other languages or build tools. Nevertheless, Maven is one of the most widely adopted infrastructures in real-world software development, making it a representative and meaningful evaluation target. Extending \toolname to support additional programming languages and build systems is a promising direction for future work to further enhance its applicability.

%% file: sections/related_work.tex
\section{Related Work}

Early research on test case updating primarily relied on program analysis techniques and heuristic rules \cite{Daniel_test_update_issta2010, Daniel_test_update_tool_icse2011, Xu_test_update_apsec2014, Mehdi_test_update_icst2012}. Daniel et al. \cite{Daniel_test_update_issta2010} proposed a symbolic-execution-based approach to update outdated tests that lack complex control flow or operations on expected values. Building on this idea, they developed the \textsc{ReAssert} tool \cite{Daniel_test_update_tool_icse2011}. \textsc{ReAssert} can automatically suggest updates to outdated unit tests, such as replacing literals, modifying assertion methods, or decomposing a single assertion into multiple assertions. Xu et al. \cite{Xu_test_update_apsec2014} formulated the insertion and deletion of method calls during test updates as a search problem and applied genetic algorithms to repair outdated test cases. 

With the advancement of machine learning and deep learning, an increasing number of studies have focused on learning-based techniques for test case updating. Shimmi et al. \cite{Shimmi_test_update_ast2022} leveraged historical test cases by embedding prior test information using neural networks and recommending updated tests through similarity matching.
Wang et al. \cite{SITAR} identified features required for test updating and proposed a machine-learning-based approach, \textsc{Sitar}, to repair outdated tests. Hu et al. \cite{CEPROT} employed Transformer-based deep learning models to update outdated test cases. Yaraghi et al. \cite{TARGET} investigated how to prioritize code change hunks relevant to test updates from a large number of change hunks and used code language models to generate updated tests. However, these studies all neglected coverage and the adequacy of test assertions.

In recent years, LLMs have demonstrated remarkable capabilities across various software engineering tasks, leading to significant progress in test case updating. Chi et al. \cite{REACCEPT} proposed \reaccept, which retrieves few-shot examples of historical test updates using retrieval-augmented generation and prompts large language models to update tests, and iteratively refining the results through dynamic execution feedback.
Xu et al. \cite{CommitUp} introduced a framework that first infers change intent, then retrieves historically similar examples, and finally applies large language models with iterative feedback to update test cases.
Note that the approaches proposed by Chi et al. \cite{REACCEPT} and Xu et al. \cite{CommitUp} focused mainly on ensuring that updated tests cover the modified code lines, whereas overlooking the ability to cover previously exercised but unchanged code lines and branches.
Liu et al. \cite{SYNTER} explored the use of language servers to retrieve information relevant to test updating, providing an initial investigation into the necessity of retrieving context information; however, their approach is limited to scenarios where the API of the method under test has changed. Zhang et al. \cite{TestUpdater} further examined information retrieval during test updating by using language servers to determine whether new dependencies need to be imported when updated tests still fail to compile or execute, reinforcing the importance of retrieving additional information, particularly in the presence of compilation or execution failures. Note that the approaches proposed by Liu et al. \cite{SYNTER} and Zhang et al. \cite{TestUpdater} both use the language server to retrieve context information. However, this retrieval method cannot handle symbols generated by LLM hallucinations or some private symbols \cite{Priyanshu_retrieval_2025}.

Our work differs from prior studies along several key dimensions. First, unlike approaches such as \reaccept~\cite{REACCEPT} and \testupdater~\cite{TestUpdater}, which require the focal method to be modified, our approach follows \target~\cite{TARGET} in prioritizing code change hunks relevant to test updates, enabling effective test case updating even when the focal method itself remains unchanged. Second, whereas prior work largely focuses on test executability and line coverage, our approach explicitly incorporates branch coverage and assertion robustness through coverage and mutation analysis, thereby improving both functional coverage and fault-detection capability of updated tests. Third, existing approaches typically rely on exact-matching–based retrieval mechanisms; in contrast, we adopt a two-stage retrieval strategy that combines semantic similarity with precise context resolution, allowing us to handle private symbols and hallucinated symbols more robustly. Finally, unlike prior studies that evaluate test updating at the commit level~\cite{REACCEPT,TARGET,TestUpdater,CommitUp}, we study test case updating at the pull-request level, which more accurately reflects real-world development workflows where code and tests often evolve across multiple commits within a single development task.
